\documentclass[twocolumn,runningheads]{svjour2}
\smartqed  
\usepackage{graphicx}
\journalname{Astrophysics and Space Science}
\begin{document}

\title{Dim Isolated Neutron Stars, Cooling and Energy Dissipation}

\titlerunning{DINs Cooling and Energy Dissipation}        
\author{M. Ali Alpar}

\institute{Sabanc{\i} University, Orhanl{\i}, Tuzla, 34956 Istanbul, Turkey\\
              Tel.: +90-216-4839510\\
              Fax: +90-216-4839550\\
\email{alpar@sabanciuniv.edu}}         
\date{Received: date / Accepted: date}
% The correct dates will be entered by the editor

\maketitle

\begin{abstract}
The cooling and reheating histories of dim isolated neutron stars(DINs) are discussed.  
Energy dissipation due to dipole spindown with ordinary and magnetar fields,
and due to torques from a fallback disk are considered as alternative 
sources of reheating which would set the temperature of the neutron star after the 
initial cooling era. Cooling or thermal ages are related to the numbers and formation 
rates of the DINs and therefore to their relations with other 
isolated neutron star populations. Interaction with a fallback disk, 
higher multipole fields and activity of the neutron star are briefly discussed.

%\PACS{First \and Second \and More}
\end{abstract}

\section{Introduction}
\label{intro}
All "magnificent" seven ROSAT detected sources display X-ray fluxes consistent with thermal luminosities of the order of 10$^{31}$ - 10$^{32}$ ergs s$^{-1}$. In particular, RX J1856.5-3754, which is the only source in this class with a parallax distance determination and kinematic age, has luminosity  3 $\times$10$^{31}$ ergs s$^{-1}$ at distance 120 pc. For RXJ1856.5-3754 and RX J 0720.4-3125 spectra can be fit with two blackbodies to cover X-ray and optical data. This is probably an indicative representation of the actual temperature modulation due to the magnetic fields. Moreover, the soft to hard luminosity ratio is the same, 0.5, in both sources, possibly indicating similar physics on the surface. The soft blackbody seems to cover the entire neutron star surface. This suggests that the entire surface luminosity is of order 10$^{31}$ - 10$^{32}$ ergs s$^{-1}$ also in the latter two sources. Are luminosities of 10$^{31}$ - 10$^{32}$ ergs s$^{-1}$ standard for these sources?\\  
\\
Cooling provides the source of thermal X-ray luminosity for a young neutron star. Cooling proceeds first via neutrino emission from the interior of the neutron star, to give way to cooling by surface photon emission at an age of the order of  10$^{6}$ yrs. According to families of standard cooling curves (eg Tsuruta et al 2002), luminosities in the range of 10$^{31}$ - 10$^{32}$ ergs s$^{-1}$ correspond to the end of the neutrino cooling era and the transition to photon cooling.\\  
\\
Why do we not  detect a few sources at age 10$^{5}$ yr and luminosity 10$^{33}$ ergs s$^{-1}$ ?
The space density of 10$^{5}$ yr old sources is only 10 times less than the density 
of the 10$^{6}$ yr old sources.  With luminosities one order of magnitude larger according to the standard cooling curves, the detectable volume may not be significantly larger than it is for the 10$^{6}$ yr old sources, also considering that the detected neighbourhood being 
limited by absorption.\\  
\\
We might expect an even larger number of older DINs, say at age 3 $\times$ 10$^{6}$ yrs. While the cooling luminosity drops sharply at ages of a few 10$^{6}$ yrs such nearby sources would be above XMM and Chandra detection limits. The lower thermal luminosities and effective temperatures of older sources means the spectra are shifted to lower photon energies and therefore detection may be more difficult because of absorption. With only 7 sources,  distance and luminosity uncertainties and lack of reliable age estimates, it is at present too early to test whether the population of DINs is consistent with being powered by cooling luminosities alone. In particular it will be interesting to see if more statistics and age determinations yield sources at ages longer than 10$^{6}$ yrs without the drop-off in luminosity expected in the photon-cooling era . With typical distances of the order of 100 pc and if their galactic birthrate is of the order of the total neutron star birthrate, 10$^{-2}$ yr$^{-1}$, the DINs may well be a population dominated by sources older than 10$^{6}$ yrs, whose luminosity is supplied by something else after the drop-off in the cooling luminosity. 

\section{Energy Dissipation}
\label{dissip}
What else can happen after the initial 10$^{6}$ yrs of cooling? 
A star with viscous coupling between different internal components has to dissipate energy when it is spun up or down by an external torque. 
Reheating of neutron stars by energy dissipation was first considered by Alpar et al (1984) and Alpar, Nandkumar \& Pines (1986) in connection with the dynamical coupling between the pinned inner crust superfluid and the crust of the neutron star. The steady state thermal luminosity
supplied by energy dissipation is of the form: 
\begin{equation}
L_{diss} = J |\dot \Omega| 
\end{equation}
where J is a parameter describing the effective viscous coupling and $\dot \Omega $ is the spindown rate due to the external torque on the 
neutron star. In the case of the pinned inner crust superfluid-vortex creep model, J$\sim$ 10$^{43}$ erg s.\\  
\\
Whether a detectable thermal luminosity can be provided by energy dissipation after the cooling luminosity drops off depends on the external torque on the star.   The external torque due to a rotating magnetic dipole gives the spindown rate
\begin{equation}
I \dot \Omega = - 2/3 \; \mu^2  \Omega ^3 / c^3
\end{equation}  
where I is the star's moment of inertia, $\mu$ is the dipole magnetic moment perpendicular to the rotation axis and $\Omega$ is the rotation rate. This yields the time dependence: 
\begin{equation}
|\dot \Omega| = 4 \times 10^{-13} s^{-2}   t_6 ^{-3/2} I_{45} ^{1/2} \mu_{30} ^{- 1}
\end{equation}  
where t$_6$ is the age in units of 10$^6$  yrs, 
I$_{45}$ the star's moment of inertia in units of 10$^{45}$ gm cm$^2$, and $\mu_{30}$ the dipole magnetic moment 
in units of 10$^{30}$ G cm$^3$.  The expected energy dissipation rate is
\begin{equation}
L_{diss} = 4 \times 10^{30} s^{-2}   t_6 ^{-3/2}  I_{45}^{1/2} \mu_{30} ^{- 1}.
\end{equation}
A nearby (d $\sim$ a few 100 pc) isolated neutron star with an "ordinary" magnetic moment $\mu_{30} \sim$ 1 might be detectable up to ages of about 10$^{7}$ yrs. For a constant dipole moment in the magnetar range, $\mu_{30} \sim$ 100-1000 the dissipation luminosity is very low, and chances of detection 
after the 10$^{6}$ yrs of initial cooling are negligible. Field decay makes the prospects even worse.\\  
\\
An alternative source of external torque on an isolated neutron star is the torque from a fallback disk. Fallback disks were proposed for anomalous X-ray pulsars by Chatterjee, Hernquist \& Narayan (2000). It was proposed by Alpar (2001) that the presence or absence, and properties, of a fallback disk is the initial condition, along with initial dipole moment and rotation rate, that determines the subsequent evolution of all classes of young neutron stars. In particular Alpar (2001) suggested that the DINs, being likely to be of ages of the order of 10$^{6}$ yrs or more, may be powered by energy dissipation due to the external torque supplied by a fallback disk (for those DINs whose cooling luminosity has already dropped off). \\  
\\
Torques from a fallback disk near rotational equilibrium with the neutron star can be estimated as
\begin{eqnarray} I |\dot \Omega | & \sim & ( \mu^2 / {r_A}^3 ) (\Omega - \Omega_{eq})/ \Omega_{eq} \nonumber \\
 & \sim & ( \mu^2 /  r_{co}^3 ) (\Omega   -\Omega_{eq} )/ \Omega_{eq} \nonumber \\
 & \sim & ( \mu^2 \Omega^2 /GM)(\Omega   -\Omega_{eq} )/ \Omega_{eq} 
\end{eqnarray}
which gives the energy dissipation rate estimate  
\begin{eqnarray}
L_{diss} & \sim & (J / I) ( \mu^2 \Omega^2 / GM) (\Omega  - \Omega_{eq} ) / \Omega_{eq} \nonumber \\
 & \sim &  10^{31} \mu_{30} ^2 \; \Omega^2 ( M/M_{sun} ) ^{-1} erg \; s^{-1},
\end{eqnarray}
taking $(\Omega   -\Omega_{eq} )/ \Omega_{eq} \sim $ 0.1 (comparable to the spread of  DIN periods). Disks surviving beyond 10$^{6}$ yrs would keep most DINs (the oldest and most abundant) at luminosities of 10$^{31}$ - 10$^{32}$ ergs s$^{-1}$ throughout the lifetime of the disk, which may be longer than 10$^{6}$ yrs.

\section{Discussion}
\label{dis}
 
At typical separations of say 150 pc, the DINs must make up a galactic plane population of about 10 000. If the duration of the neutrino cooling epoch with L larger than 10$^{32}$ ergs s$^{-1}$ is taken to be their representative age, this age is 10$^6$ yrs. The rate of formation is therefore of the order of 10$^{-2}$ yr$^{-1}$ .  The DINs make up a substantial fraction of the supernova rate: they are a very abundant population of young neutron stars, comparable to isolated radio pulsars (and possibly RRATS).\\  
\\
Dipole magnetic fields inferred from $\dot{P}$ and P, as well as surface fields inferred from absorption features from proton cyclotron lines are about 6 $\times$10$^{13}$ G. This is of the order of or above the quantum critical field, and only an order of magnitude less than the inferred dipole magnetic fields of AXPs and SGRs.  Periods are clustered in the same special narrow range as AXPs and SGRs, for 5 DINs observed periods range from 3.45 s to 11.37 s (Haberl 2005). This period clustering can be explained with fallback disks acting as a gyrostat around the neutron star, and determining its asymptotic equilibrium period (Alpar 2001). In this scheme DINs, AXPs and SGRs are supposed to be in a propeller interaction with the fallback disk.   Most numerous, after the no-disk radio pulsars, must be the very small mass disk cases, where the disk would have a long lifetime. These are proposed to be the DINs.\\  
\\
All these classes of objects are in the upper right hand part of the P-$\dot{P}$ diagram. All are close to but above the death line for radio pulsars. They share the P-$\dot{P}$ neighbourhood with the high dipole field radio pulsars. So why do the DINs (AXPs and SGRs also) not function as radio pulsars? This is more of a question for the DINs as their inferred dipole surface fields are of the same order as the highest dipole field radio pulsars: the AXPs and SGRs might short out the pair creation in the polar cap more easily by virtue of fields that are one order of magnitude larger than those of the radio pulsars. 
Now there is a new enigma: RRATS (McLaughlin et al 2006), some of which are close to same corner of P-Pdot diagram. RRATS also may be similar to AXPs and SGRs in that they burst, though in the radio.
What distinguishes between the different classes might indeed be the presence or absence, and nature of a fallback disk around the neutron star.  The $\dot{M}$ history of the disk determines the equilibrium period to which the star approaches asymptotically.  For B = 10$^{12}$ - 10$^{13}$ G ,  a relatively wide range of $\dot{M}$ gives a narrow range of equilibrium periods:   P$_{eq}$ scales with $ |\dot{M}|^{- 4/7}$. An interesting example is that  if the dipole magnetic field on the surface is the quantum critical field, and $\dot{M}$ is at the Eddington value, this leads to equilibrium periods of about 12 s. \\  
\\
This approach is not in conflict with magnetar models (Thompson \& Duncan 1995, Woods \& Thompson 2006).
The magnetar model functions on the basis of very strong magnetic fields anchored in the neutron star crust. These fields are likely to develop and release energy in local processes in the neutron star crust. The surface magnetar fields are likely to be concentrated in the higher multipoles, and the dipole magnetic field on the surface of magnetars may be of order or less than the quantum critical field? One might speculate that when the quantum critical field is reached on the surface, leading to pair creation, near surface currents and surface heating, magnetic structure in the higher multipoles will be amplified, at the expense of the dipole (global) surface field, which might be limited to a value near the quantum critical field. Arguments for surface multipole fields have been brought up in several connections (Zane \& Turolla 2005; see also the contributions by Zane, by Tiengo and by Gaensler at this conference). \\
\\
In the context of fallback disk models and hybrid models with magnetar fields in the higher multipoles, the presence of fallback disk torques interacting with the dipole component of the magnetic field would lead to detectable thermal luminosities from the DINs if the disk lifetime extends beyond the 10$^6$ yr timescale of the cooling luminosity. As explored here, this is because the energy dissipation rate for propeller disk torques near equilibrium is stronger than energy dissipation rates supplied by ordinary or magnetar dipole spindown, and may be roughly constant throughout the disk lifetime, unlike the dipole spindown case for which the energy dissipation rate decays rapidly with a power law time dependence.\\

\begin{acknowledgements}
I thank the Sabanc{\i} University Astrophysics and Space Forum and the Turkish Academy of Sciences for research support.
\end{acknowledgements}

\end{document}